\documentclass[twoside,12pt]{article}
\usepackage{epsfig}

\newcommand{\ba}[1]{\begin{eqnarray} \label{(#1)}}
\newcommand{\ea}{\end{eqnarray}}

\newcommand{\AmS}{{\protect\the\textfont2
  A\kern-.1667em\lower.5ex\hbox{M}\kern-.125emS}}

\def\be{\begin{equation}}
\def\ee{\end{equation}}
\def\bea{\begin{eqnarray}}
\def\eea{\end{eqnarray}}

\begin{document}

\title{Search for Cold Dark Matter and Solar Neutrinos 
	with GENIUS and GENIUS-TF}

\author{I.V. Krivosheina
\\
Radiophysical Research Institute (NIRFI),
	Nishnii-Novgorod, Russia,
\\ 
	Max-Planck-Institut f\"ur Kernphysik,
	P.O. Box 10 39 80, D-69029 Heidelberg,
\\ 
	Germany, E-mail: irina@gustav.mpi-hd.mpg}

\maketitle


\begin{abstract}
	The new project 
	GENIUS will cover a wide range of the parameter 
	 space of predictions of SUSY for neutralinos as cold dark matter. 
	 Further it has the potential to be a real-time detector 
	 for low-energy ($pp$ and $^7$Be) solar neutrinos. 
	 A GENIUS Test Facility has just been funded and will 
	 come into operation by end of 2002.
\end{abstract}

\section{Introduction}

	Concerning solar neutrino physics, present information on 
	possible $\nu$ oscillations relies on 0.2\% of the solar neutrino flux.
	The total $pp$ neutrino flux has not been measured and also no 
	real-time information is available for the latter.
	Concerning the search for cold dark matter, direct detection of 
	the latter by underground detectors remains 
	indispensable. 
	
	The GENIUS project proposed in 1997  
\cite{KK-Bey97,GEN-prop,KK60Y,KK-InJModPh98,KK-J-PhysG98} 
	as the first third generation $\beta\beta$ detector, 
	could attack all of these problems with an unprecedented sensitivity. 
	 GENIUS will allow real time detection 
	of low-energy solar neutrinos with a threshold of 19 keV.
	   For the further potential of GENIUS for other beyond 
	SM physics, such 
	as double beta decay, SUSY, compositeness, leptoquarks, 
	   violation of Lorentz invariance and equivalence principle, 
	   etc we refer to  
\cite{KK-Erice01,KK-SprTracts00,KK-InJModPh98,KK60Y,KK-WEIN98,KK-Neutr98}.

                      
\begin{figure}[t]
\epsfig{file=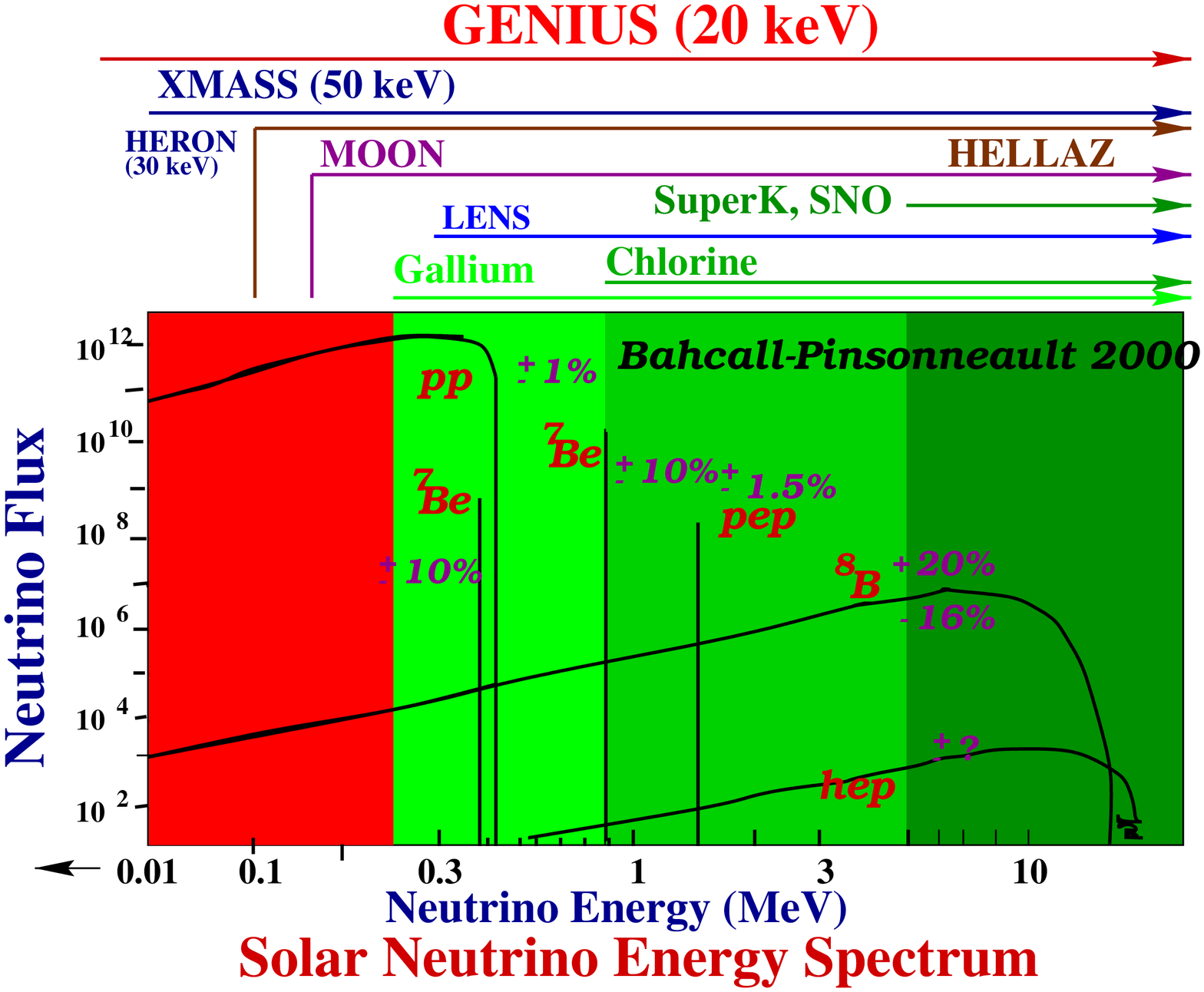,scale=0.4}
\hspace{.5cm}
\includegraphics*[scale=0.45]
{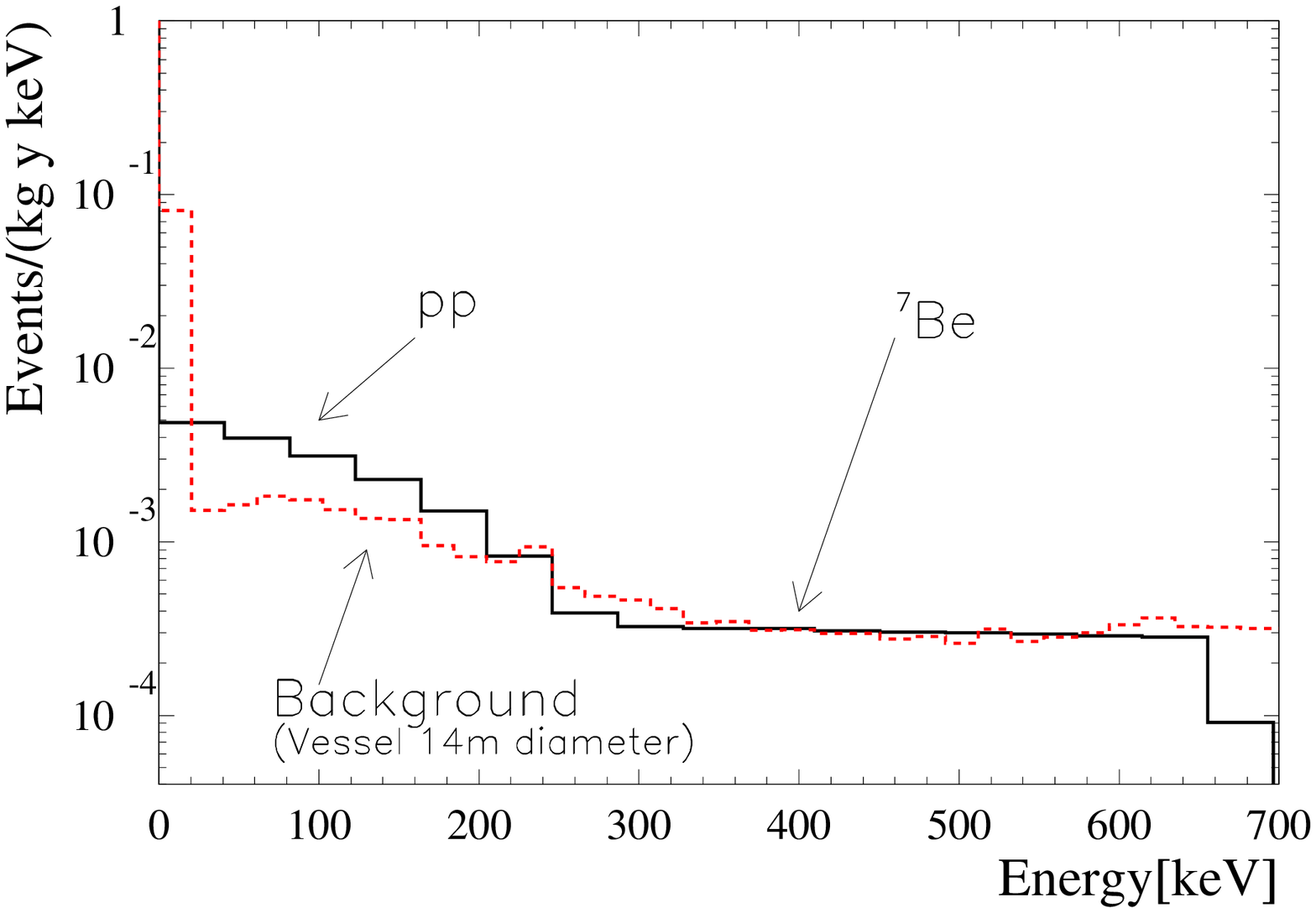}
\caption[]{
       Left: 
	The sensitivity (thresholds) of different running and projected 
	solar neutrino detectors (see 
\protect\cite{HomP-Bach} and home-page
	HEIDELBERG NON-ACCELERATOR PARTICLE PHYSICS 
	GROUP: $http://www.mpi-hd.mpg.de/non\_acc/$). 
	Right:
	Simulated spectrum of low-energy solar neutrinos 
       (according to SSM) for the GENIUS detector 
       (1 tonne of Ge) (from  
\protect\cite{KK01}, 
	and estimated background).
\label{fig:sol-neutr-Bach}}
\end{figure}


\section{GENIUS and Low-Energy Solar Neutrinos}

	GENIUS which has been proposed for solar $\nu$ detection 
	in 1999  
\cite{BKK-SolN,GEN-prop}
	, could be the first detector measuring 
		the {\em full}\ $pp$ (and $^7$Be) 
		neutrino flux in real time 
(Fig.~\ref{fig:sol-neutr-Bach}).

	The main idea of GENIUS, originally proposed for double beta and dark 
	matter search 
\cite{KK-Bey97,KK-J-PhysG98,KK-Neutr98,KK-WEIN98,KK-InJModPh98,KK-SprTracts00}
  is to achieve an extremely low radioactive background 
	(factor of $>$  1000 smaller than in the HEIDELBERG-MOSCOW 
	experiment) by using 'naked' detectors in liquid nitrogen.

	While for cold 
	dark matter search 100 kg of {\it natural} Ge detectors 
	are sufficient, GENIUS as a solar neutrino detector would contain 
	1-10 tons of enriched $^{70}{Ge}$ or $^{73}{Ge}$.

	That Ge detectors in liquid nitrogen operate excellently, has been 
	demonstrated in the Heidelberg low-level laboratory  
\cite{KK-J-PhysG98,Bau98}
	and the overall feasibility of the project has been shown in 
\cite{GEN-prop,KK-J-PhysG98,KK-NOW00,LowNu2}.

	The potential of GENIUS to measure the spectrum of low-energy solar 
	neutrinos in real time has been studied by  
\cite{BKK-SolN,GEN-prop,LowNu2}. 
	The detection reaction is elastic neutrino-electon scattering 
	$\nu ~+~ e^- \longrightarrow~ \nu~ +~e^-$. 

	The maximum electron recoil energy is 261 keV for the pp neutrinos 
	and 665 keV for the $^{7}{Be}$ neutrinos. 
	The recoil electrons can be detected through their ionization 
	in a HP Ge detector with an energy resolution of 0.3$\%$. GENIUS 
	can measure only (like BOREXINO, and others) but with much better 
	energy resolution) the energy distribution of the recoiling electrons, 
	and not directly determine the energy of the incoming neutrinos. 
	The dominant part of the signal in GENIUS is produced by $pp$ 
	neutrinos (66$\%$) and $^{7}{Be}$ neutrinos (33$\%$). The detection 
	rates for the $pp$ and $^{7}{Be}$ fluxes are according to the 
	Standard Solar Model  
\cite{BacBasPins98}
	$R_{pp}$ = 35 SNU = 1.8  events/day ton~ (18 events/day 10 tons) 
	and~ $R_{^{7}{Be}}$~ = 13 SNU~ = 0.6 events/day ton~ 
	(6 events/day 10 tons)~(1 SNU = ${10}^{-36}$/s target atom).

	To measure the low-energy solar $\nu$ flux with a signal to 
	background ratio of 3:1, the required background rate is 
	about 1 $\times~ {10}^{-3}$ events/kg y keV in this energy range. 
	This is about a factor of 10 smaller than what is required for 
	the application of GENIUS for cold dark matter search. This can 
	be achieved if the liquid nitrogen shielding is increased to at 
	least 13 m in diameter and production of the Ge detectors is 
	performed underground (see 
\cite{BKK-SolN,LowNu2}).

	Another source of background 
	is coming from   
	2$\nu\beta\beta$ decay of $^{76}{Ge}$, which is contained in 
	{\it natural} Ge with 7.8$\%$. Using enriched 
	$^{70}{Ge}$ or $^{73}{Ge}$ ($>$85$\%$) 
	as detector material, the abundance of the $\beta\beta$ emitter 
	can be reduced up to a factor of 1500. In this case the $pp$-signal 
	will not be disturbed by 2$\nu\beta\beta$ decay (see 
\cite{LowNu2}). 

	The expected spectrum of the low-energy signal in the SSM is 
	shown in 
Fig. ~\ref{fig:sol-neutr-Bach} (right part).  

	After the unfavouring of the SMA solution by 
	Superkamiokande, it is important to 
	differentiate between the LMA and the LOW solution. Here 
	due to its relatively high counting rate, GENIUS will be able to test  
	in particular the LOW solution of the solar $\nu$ problem by the 
	expected day/night variation of the flux (see 
\cite{KK-NOW00,LowNu2}).



\begin{figure}
\begin{center}
\epsfig{file=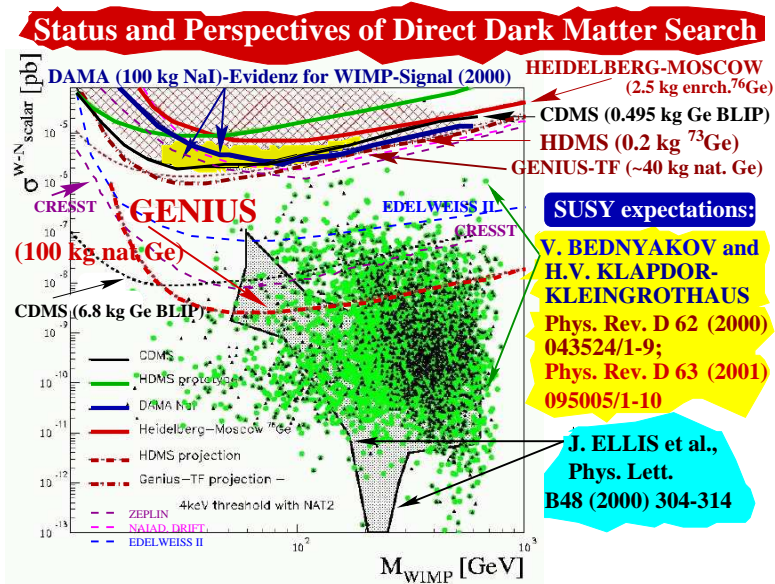,scale=0.4}
\end{center}
\caption[]{
       WIMP-nucleon cross section limits in pb for scalar interactions as 
       function of the WIMP mass in GeV. 
       Shown are contour lines of present experimental limits (solid lines) 
       and of projected experiments (dashed lines). 
       Also shown is the region of evidence published by DAMA. 
       The theoretical expectations are shown by two scatter plots, 
	- for accelerating and for non-accelerating Universe (from  
\cite{BedKK00,BedKK-01}) and by the grey region (from  
\cite{EllOliv-DM00}). 
	{\em Only}\ GENIUS will be able to probe the shown range 
       also by the signature from seasonal modulations.
\label{fig:Bedn-Wp2000}}
\end{figure}


\section{GENIUS and Cold Dark Matter Search}
 
	GENIUS would already in a first step, with 100 kg of 
		{\it natural} Ge detectors, cover a significant part of the 
		MSSM parameter space for prediction of neutralinos 
		as cold dark matter 
(Fig.~\ref{fig:Bedn-Wp2000}) 
	(see, e.g. 
\cite{BedKK00,BedKK-01,EllOliv-DM00})
	For this purpose the background in the energy range 
		$< 100$~keV has to be reduced to 
		$10^{-2}$ (events/kg y keV). 
	At this level solar neutrinos as source of background 
	are still negligible. 
Fig.~\ref{fig:Bedn-Wp2000} 
	shows together with the expected sensitivity of GENIUS, 
	for this background, predictions for neutralinos as dark matter by 
	two models, one basing on supergravity  
\cite{EllOliv-DM00}, another basing on the MSSM with more 
     relaxed unification conditions  
\cite{BedKK00,BedKK-01}.

	     The sensitivity of GENIUS for Dark Matter corresponds to 
	     that obtainable with a 1 km$^3$ AMANDA detector for 
	     {\it indirect} detection (neutrinos from annihilation 
	     of neutralinos captured at the Sun) (see  
\cite{Eds99}). 
	Interestingly both experiments would probe different neutralino 
	compositions: GENIUS mainly gaugino-dominated neutralinos, 
	AMANDA mainly neutralinos with comparable gaugino and 
	Higgsino components (see Fig. 38 in  
\cite{Eds99}). 
     It should be stressed that, together with DAMA, GENIUS will be 
     {\em the only}\ future Dark Matter experiment, which would be able to 
     positively identify a dark matter signal by the seasonal 
     modulation signature. 
     This {\it cannot} be achieved, for example, by the CDMS experiment.

\section{GENIUS-TF}

	As a first step of GENIUS, a small test facility, GENIUS-TF, 
	is at present under installation in the Gran Sasso 
	Underground Laboratory 
\cite{GENIUS-TF}.
	With about 40 kg of natural Ge detectors operated 
	in liquid nitrogen, GENIUS-TF could test the DAMA seasonal 
	modulation signature for dark matter. 
	No other experiment running, like CDMS, IGEX, etc., 
	or projected at present, will have this potential 
\cite{KK-LP01}.
	Up to summer 2001, already six 2.5 kg Germanium detectors with 
	an extreme low-level threshold of $\sim$500 eV have been produced.
	



\begin{thebibliography}{99}




\bibitem{KK60Y}
		H.V. Klapdor-Kleingrothaus, 
		{\sf "60 Years of Double Beta Decay"}, {\it World Scientific, 
		Singapore} (2001) 1253~p.


\bibitem{Bau98}
 	L.~Baudis, G.~Heusser, B.~Majorovits, Y.~Ramachers, H.~Strecker and 
 	H.V.~Klapdor--Kleingrothaus, {\it hep-ex/} {\bf 9811040} and 
	{\it Nucl. Instr. Meth.} {\bf A 426}, 425 (1999).


\bibitem{BKK-SolN}
	L. Baudis and H.V. Klapdor-Kleingrothaus, 
	{\it Eur. Phys. J.} {\bf A 5}, 441-443  (1999) and in 
	Proceedings of the 2nd Int. Conf. on Particle 
	Physics Beyond the Standard Model BEYOND'99, 
	Castle Ringberg, Germany, 6-12 June 1999, 
	edited by H.V. Klapdor-Kleingrothaus and I.V. Krivosheina, 
	{\it IOP Bristol}, 1023 - 1036 (2000).

\bibitem{KK01}
		H.V. Klapdor-Kleingrothaus et al., 
		to be publ. 2001 
		and ${\it http://www.mpi-hd.mpg.de/non\_acc/}$


\bibitem{KK-LP01}
	H.V. Klapdor-Kleingrothaus, 
	in Proc. of the XX Lepton Photon Symposium (LP01), 
	July 23 - 28, 2001, Rome, Italy, 
	{\it World Scientific, Singapore} (2002) 

\bibitem{KK-Erice01}
	H.V. Klapdor-Kleingrothaus, 
	in Proc. of the Erice School on Nuclear Physics about Neutrinos, 
	ed. A. Faessler,  
	{\it Progress in Particle and Nuclear Physics} {\bf 48} (2002).

 	
\bibitem{KK-Bey97}
	H.V. Klapdor-Kleingrothaus in Proceedings of BEYOND'97, 
	First International Conference on Particle Physics 
	Beyond the Standard Model, Castle Ringberg, Germany, 
	8-14 June 1997, 
     edited by H.V. Klapdor-Kleingrothaus and H. P\"as, 
	{\it IOP Bristol} 485-531 (1998)  


\bibitem{GEN-prop}
	H.V. Klapdor-Kleingrothaus et al. 
	{\it MPI-Report} {\bf MPI-H-V26-1999} and 
	{\it Preprint: hep-ph/}{\bf 9910205} and in 
	Proceedings of the 2nd Int. Conf. on Particle 
	Physics Beyond the Standard Model BEYOND'99, 
	Castle Ringberg, Germany, 6-12 June 1999, 
	edited by H.V. Klapdor-Kleingrothaus and I.V. Krivosheina, 
	{\it IOP Bristol}, 915 - 1014 (2000).

\bibitem{KK-Neutr98}
	H.V. Klapdor-Kleingrothaus, in Proc. of 18th Int. 
	Conf. on Neutrino Physics and Astrophysics (NEUTRINO 98), 
	Takayama, Japan, 4-9 Jun 1998, (eds) Y. Suzuki et al. 
	{\it Nucl. Phys. Proc. Suppl.} {\bf 77}, 357 - 368 (1999).   


\bibitem{KK-WEIN98}
	H.V. Klapdor-Kleingrothaus, in Proc. of WEIN'98, 
	"Physics Beyond the Standard Model", Proceedings of the Fifth Intern.  
	WEIN Conference, P. Herczeg, C.M. Hoffman and 
	H.V. Klapdor-Kleingrothaus (Editors), 
	{\it World Scientific, Singapore}, 275-311 (1999).



\bibitem{KK-NOW00}
	H.V. Klapdor-Kleingrothaus, in Proc. of Int. Conference 
	NOW2000 - "Origins of Neutrino Oscillations", 
	{\it Nucl. Phys.} {\bf B} (2001) ed. G. Fogli and 
	{\it Preprint: hep-ph/}{\bf 0102277}, 
	{\it Preprint: hep-ph/}{\bf 0102276}. 



\bibitem{LowNu2}
	H.V. Klapdor-Kleingrothaus, in 
	Proc. Int. Workshop on Low Energy Solar Neutrinos, LowNu2, 
	December 4 and 5 (2000) Tokyo, Japan, 
	ed: Y. Suzuki, World Scientific, Singapore (2001), home page: 
	{\it http://www-sk.icrr.u-tokyo.ac.jp/neutlowe/2/transparency/
index.html}


\bibitem{KK-J-PhysG98}
	H.V. Klapdor-Kleingrothaus, J. Hellmig and M. Hirsch, 
	{\it J. Phys.} {\bf G 24}, 483 (1998).

\bibitem{KK-InJModPh98}
	H.V. Klapdor-Kleingrothaus, {\it Int. J. Mod. Phys.} {\bf A 13}, 3953  
	(1998).

\bibitem{KK-SprTracts00}
	H.V. Klapdor-Kleingrothaus, {\it Springer Tracts in Modern Physics}, 
	{\bf 163}, 69-104 (2000), 
	{\it Springer-Verlag, Heidelberg, Germany} (2000).



\bibitem{BedKK00}
	V.A. Bednyakov and H.V. Klapdor-Kleingrothaus, 
	{\it Phys. Rev.} {\bf D 62} (2000) 043524/1-9 and {\it hep-ph/}
	{\bf 9908427}. 

\bibitem{BedKK-01}
	V.A. Bednyakov and H.V. Klapdor-Kleingrothaus, 
	{\it Preprint: hep-ph/}{\bf 0011233} (2000) and  
	{\it Phys. Rev.} {\bf D 63} (2001) 095005.

\bibitem{EllOliv-DM00}
	J. Ellis, A. Ferstl and K.A. Olive, {\it Phys. Lett.} {\bf B 481},  
	304--314 (2000) and {\it Preprint: hep-ph/}{\bf 0001005} 
	and {\it Preprint: hep-ph/}{\bf 0007113}.

\bibitem{Eds99}
	J. Edsj\"o, Neutralinos as dark matter - can we see them? 
	Seminar given in the theory group, Department of Physics, 
	Stockholm University, October 12, 1999, home page: 
	{\it http://www.physto.se/edsjo/}

\bibitem{GENIUS-TF}
	H.V. Klapdor-Kleingrothaus, L. Baudis, A. Dietz, 
	G. Heusser, I.V. Krivosheina, B. Majorovits and 
	H. Strecker, hep-ex/0012022, Subm. for Publ. (2001).	


\bibitem{HomP-Bach}
	see: {\it http://www.sns.ias.edu.jnb/}

\bibitem{Bach89}
	J.N. Bahcall, {\it Neutrino Astrophysics}, Cambridge Univ. 
	Press (1989). 

\bibitem{BacBasPins98}
	J.N. Bahcall, S. Basu and M. Pinsonneault, 
	{\it Phys. Lett.} {\bf B 433}, 1 (1998).


\end{thebibliography}
\end{document}